\documentclass[floats,floatfix,showpacs,amssymb,prd,twocolumn,superscriptaddress,nofootinbib]{revtex4-1}

\usepackage{amssymb,amsmath,verbatim,mathtools,needspace,enumitem,etoolbox,graphicx}

\usepackage[dvipsnames, usenames]{xcolor}

\definecolor{linkcolor}{rgb}{0.0,0.3,0.5}
\usepackage[unicode, colorlinks=true, linkcolor=linkcolor, citecolor=linkcolor, filecolor=linkcolor,urlcolor=linkcolor, pdfusetitle]{hyperref}
\usepackage[all]{hypcap}
\usepackage[T1]{fontenc}
\usepackage[utf8]{inputenc}

\usepackage{epstopdf}

\usepackage{bm}
\usepackage{dcolumn}
\usepackage{longtable}
\usepackage{enumerate}
 \usepackage[normalem]{ulem}

\def\be{\begin{equation}}
\def\ee{\end{equation}}
\newcommand{\beq}{\begin{eqnarray}}
\newcommand{\eeq}{\end{eqnarray}}

\begin{document}

\title{Parametrized black hole quasinormal ringdown. \\I. Decoupled equations for nonrotating black holes}
\author{Vitor Cardoso}
\affiliation{CENTRA, Departamento de F\'{\i}sica, Instituto Superior T\'ecnico -- IST, Universidade de Lisboa -- UL,
Avenida Rovisco Pais 1, 1049 Lisboa, Portugal}
\affiliation{Theoretical Physics Department, CERN 1 Esplanade des Particules, Geneva 23, CH-1211, Switzerland}

\author{Masashi Kimura}
\affiliation{CENTRA, Departamento de F\'{\i}sica, Instituto Superior T\'ecnico -- IST, Universidade de Lisboa -- UL,
Avenida Rovisco Pais 1, 1049 Lisboa, Portugal}

\affiliation{Department of Physics, Rikkyo University, Tokyo 171-8501, Japan}

\author{Andrea Maselli}
\affiliation{Dipartimento di Fisica, ``Sapienza'' Universit\`a di Roma \& Sezione INFN Roma1, P.A. Moro 5, 00185, Roma, Italy}

\author{Emanuele Berti}
\affiliation{Department of Physics and Astronomy, Johns Hopkins University, 3400 N. Charles Street, Baltimore, MD 21218, US}

\author{Caio F. B. Macedo}
\affiliation{Campus Salin\'opolis, Universidade Federal do Par\'a, Salin\'opolis, Par\'a, 68721-000 Brazil}

\author{Ryan McManus}
\affiliation{Department of Physics and Astronomy, Johns Hopkins University, 3400 N. Charles Street, Baltimore, MD 21218, US}

\begin{abstract}
Black hole solutions in general relativity are simple. The frequency spectrum of linear perturbations around these solutions (i.e., the quasinormal modes) is also simple, and therefore it is a prime target for fundamental tests of black hole spacetimes and of the underlying theory of gravity. The following technical calculations must be performed to understand the imprints of any modified gravity theory on the spectrum: 1. Identify a healthy theory; 2. Find black hole solutions within the theory; 3. Compute the equations governing linearized perturbations around the black hole spacetime; 4. Solve these equations to compute the characteristic quasinormal modes. In this work (the first of a series) we assume that the background spacetime has spherical symmetry, that the relevant physics is always close to general relativity, and that there is no coupling between the perturbation equations.
Under these assumptions, we provide the general numerical solution to step 4. We provide publicly available data files such that the quasinormal modes of {\em any} spherically symmetric spacetime can be computed (in principle) to arbitrary precision once the linearized perturbation equations are known. We show that the isospectrality between the even- and odd-parity quasinormal mode spectra is fragile, and we identify the necessary conditions to preserve it. Finally, we point out that new modes can appear in the spectrum even in setups that are perturbatively close to general relativity.
\end{abstract}

\date{\today}
\maketitle

\section{Introduction}\label{sec:intro}
The coalescence of two black holes (BHs) is a fascinating process from a theoretical perspective: two regions devoid of matter interact violently to produce enormous amounts of gravitational radiation, leaving behind a stationary, vacuum solution of the field equations. In general relativity (GR), this process is particularly interesting since both the initial BH spacetimes and the final state are characterized by only two parameters: their mass and angular momentum~\cite{Chrusciel:2012jk,Cardoso:2016ryw}.

The transition from the initial to the final state is described by the so-called ``ringdown'' phase, during which the distorted remnant sheds away the nontrivial multipolar structure through gravitational waves, relaxing to the final stationary solution. 
This exponential approach to equilibrium is a property of many physical systems. It can be mathematically described as a small deviation away from the {\it final} state, or as the time window in which the spacetime response to generic perturbations can be decomposed into a set of quasinormal modes (QNMs) characterized by complex frequencies~\cite{Chandrasekhar:1975zza,Berti:2009kk}.
The ringdown is of special relevance in the context of BH physics: if GR is correct, then QNM frequencies must depend only on the BH mass and spin, and therefore ringdown observations can provide simple and clean tests of GR~\cite{Berti:2005ys,Berti:2016lat,Meidam:2014jpa,Berti:2015itd,TheLIGOScientific:2016src,Berti:2018vdi}.

It is generally accepted that GR does not offer the ultimate description of gravity. Rather, discrepancies in its strong-field or large-scale descriptions will eventually become apparent in a manner consistent with some of its known conceptual problems~\cite{Berti:2015itd,Barack:2018yly}.
Thus, one expects that BH solutions will differ, if only slightly, from those of GR, and that the dynamics of BHs will also be described by slightly different equations.
The correct description of gravity at all scales and energies is presently unknown. 
Traditionally, the strategy has been to identify modified theories of gravity which contain new features in their dynamics that could be smoking guns of new physics. 
Once the theory is identified, one computes its BH solutions, studies linearized perturbations around these BH solutions, and finally solves the relevant equations to compute QNM frequencies. 

Recent efforts have focused instead on understanding the possible {\em functional form} of such corrections, using resummation methods. 
For example, effective field theory (EFT) arguments can produce the general form of the corrections in the absence of extra fundamental fields in the problem~\cite{Endlich:2017tqa}. Such corrections can then be used to study BH solutions and their dynamics, most notably their QNM spectra~\cite{Cardoso:2018ptl}. 
Once new fundamental fields -- for example scalars -- are allowed, the dynamics become intrinsically more complicated and richer phenomenology arises from, for instance, the coupling between the scalar and the gravitational degrees of freedom~\cite{Tattersall:2017erk,Tattersall:2018nve,Franciolini:2018uyq}. 

The (technical) burden of computing the QNMs cannot be avoided. 
In the majority of the literature, tools yielding approximate answers, such as WKB techniques~\cite{Schutz:1985,Iyer:1987}, have been used. The errors involved in the WKB approximation can be rather large even in GR (see e.g. Fig.~3 of Ref.~\cite{Berti:2009kk}), and they are poorly known in modified theories of gravity~\cite{Glampedakis:2017dvb}.
Clearly, when testing GR precision physics will be involved. 
In addition, it quickly becomes apparent that one is repeatedly solving the same equation (or systems of equations) with the same boundary conditions and with results (the QNM frequencies) which are but small modifications to the GR predictions. 
This work is intended as a first step in the general program to addressing the repetitive task of finding the eigenvalues (i.e., the QNM frequencies) of these ordinary differential equations. 
Once the linearized equations of motion around a BH spacetime are known, our results can be used to find the corresponding QNM frequencies. 
In this first work, for simplicity, we assume that there are no couplings between the perturbation equations. 
These complications will be addressed in a forthcoming publication~\cite{McManus:2019}.

We also make the rather restrictive working hypothesis that the underlying equations are separable, and that the background GR solution is nonspinning. 
Many known modified theories of gravity indeed yield separable equations for nonspinning background spacetimes, but it is unclear whether this feature is generic. 
Certainly BHs in our universe are spinning, generically breaking this hypothesis. 
Nevertheless, our results are interesting on their own and provide a better understanding of GR itself. Our formalism is theory-agnostic and background-agnostic. While in principle it only applies to nonspinning backgrounds, in Sec.~\ref{subsec:Kerr} we will show that, in fact, it can be used to accommodate (in a perturbative sense) the QNM frequencies of slowly spinning Kerr BHs. Therefore our results seem to apply more broadly than these restrictions might imply.

The plan of the paper is as follows. In Sec.~\ref{sec:framework} we present the general perturbative framework underlying our analysis.
In Sec.~\ref{sec:results} we present our main results and their implications for isospectrality. 
In Sec.~\ref{sec:asymptotics} we discuss analytically and numerically the convergence of the perturbative expansion, and we present a simple example of a perturbation to the known GR potential that may lead to divergences.
In Sec.~\ref{sec:examples} we verify our formalism in some interesting cases where QNM solutions are known: an EFT calculation of deviations from GR, the Reissner-Nordstr\"om (RN) solution, and scalar perturbations are slowly rotating BHs.
In Sec.~\ref{sec:errors} we present a preliminary estimate of the conditions under which perturbative deviations from the GR frequencies could be detectable.
We conclude in Sec.~\ref{sec:discussion} with a discussion of the limitations of this analysis and directions for future work.
In Appendix~\ref{app:transformation} we prove that our assumed form for the perturbative potentials is general enough for our present purposes, and in Appendix~\ref{app:eikonal} we study the large-$\ell$ (eikonal) limit of our approximation.

\section{Framework}\label{sec:framework}
In GR, gravitational fluctuations have two degrees of freedom, which can be described by a linear combination $\Phi$ of the metric perturbations. In a nonspinning background, such modes are described by an ``axial'' (or odd, or Regge-Wheeler) and a ``polar'' (or even, or Zerilli) equation of the form~\cite{Regge:1957td,Zerilli:1970se},
\be
f \frac{d}{dr}\left(f \frac{d\Phi}{dr}\right) +[\omega^2 - f V_\pm] \Phi = 0\,,\label{master}
\ee
where $f =1-r_H/r$, and $r_H$ is the horizon radius in standard Schwarzschild coordinates.

The effective potential $V_\pm$ is
\begin{equation}
  V_- = \frac{\ell(\ell + 1)}{r^2} - \frac{3 r_H}{r^3}
  \label{VGRm}
\end{equation}
for odd perturbations, and
\be
V_+ =
\frac{9\lambda r_H^2 r+3\lambda^2 r_H r^2 +  \lambda^2(\lambda+2) r^3+9 r_H^3}{r^3 (\lambda r+3 r_H)^2}
\label{VGRplus}
\ee
for even perturbations, where $\lambda =\ell^2+\ell-2$, and $\ell$ is an angular harmonic index labeling the tensorial spherical harmonics used to separate the angular coordinates from the problem (the azimuthal index can be set to $m=0$ for spherically symmetric backgrounds).

The dynamics of test scalar or vector fields in GR is also described by similar master equations, with potentials~\cite{Berti:2009kk}
\be
V_{s}=\frac{\ell(\ell + 1)}{r^2} + (1-s^2)\frac{r_H}{r^3}\,,
\ee
where $s=0,\,1$ for scalar and vector perturbations, respectively.

Given the strong observational constraints on deviations from GR~\cite{Will:2014kxa}, it might be expected that a more fundamental description of gravity will add small corrections to the GR description of astrophysical BHs. 
Such corrections will slightly disturb both the equilibrium configurations (the Schwarzschild geometry) {\it and} the equations governing dynamical fluctuations~\cite{Barausse:2008xv}. Our guiding light is asymptotically flat spacetime and, therefore, we parametrize the modified master equations by adding power-law corrections to the effective potential:
\beq
V&=&V_\pm+\delta V_\pm\,,\label{expansionV1}\\
\delta V_\pm&=& \frac{1}{r_H^2}\sum_{j=0}^\infty \alpha^\pm_j \left(\frac{r_H}{r}\right)^j
\label{expansionV}
\eeq
for gravitational fluctuations and 
\beq
V&=&V_{s}+\delta V_{s}\,,\\
\delta V_{s}&=&\frac{1}{r_H^2}\sum_{j=0}^\infty \beta^{s}_j \left(\frac{r_H}{r}\right)^j
\label{expansionVS}
\eeq
for scalars and vectors. Here, $\alpha^\pm_j, \beta^{s}_j$ are constant coefficients. Since the maximum value of the $j$-th term $ f(r)\alpha_j^\pm (r_H/r)^j$ is $\alpha_j^\pm (1+1/j)^{-j} /(j+1)$, the smallness of the corrections translates to the criterion:
\be
(\alpha_j^\pm, \beta_j^s) \ll (1+1/j)^j (j+1)\ .
\ee

Note that $\alpha^\pm_0$ and $\beta^s_0$ in Eqs.~\eqref{expansionV} and~\eqref{expansionVS} are effectively mass-square terms, hence they should be positive. Similar constraints may apply to the $j=1$ terms, which dominate the potential at large distances and which are therefore constrained by weak-field physics. For example, a $1/r$ term appears when dealing with charged scalars in a {\it charged} background (see, e.g., Eq.~(4) in Ref.~\cite{Hod:1997mt} or Eq.~(4.54) in Ref.~\cite{Brito:2015oca}). As we show in Sec.~\ref{sec:examples}, the leading-order coefficients ($\alpha_0, \beta_0$ and $\beta_1$, in the notation of this section) are nonzero for charged (RN) or slowly rotating Kerr BHs. Although these terms are dominant in the power-law expansion -- falling off even slower than the GR terms -- they should still be considered small when compared to the GR potential $V_{\pm,s}$, due to the smallness of the parameters $\alpha^\pm_j,\beta^s_j$.

In general, perturbation equations can be reduced to the form \eqref{master}, but with $f\neq 1-r_H/r$. In Appendix~\ref{app_wavereduction} we prove this for scalar and vector perturbations of static, spherically symmetric spacetimes, which indeed obey a generalized wave equation [Eq.~\eqref{generalmastereq}]. However in Appendix~\ref{app:transformation} we show that, at linear order in perturbations, Eq.~\eqref{generalmastereq} is in fact equivalent to Eq.~\eqref{master}. Gravitational perturbations depend on the chosen modified gravity theory, so it is nontrivial to provide a general proof that they always reduce to Eq.~\eqref{generalmastereq}, but in Appendix~\ref{app_wavereduction} we provide arguments supporting this conclusion. Note that, as we show below, even perturbations of slowly rotating BHs can be accommodated within our formalism, so its regime of validity may be larger than anticipated.

The parametrization used here embodies both the change in the background spacetime and the change in the dynamical equations.
It assumes no couplings to other fundamental degrees of freedom, and it also assumes no coupling between the odd and even modes; these issues will be addressed in a forthcoming study~\cite{McManus:2019}.

As predicted by perturbation theory~\cite{Leung:1999rh,Barausse:2014tra}, we find that the QNM frequencies for these potentials are described by small corrections to the GR modes~\cite{Leaver:1985ax,Berti:2005ys,Berti:2009kk}:
\beq
\omega^{\pm}_{\rm QNM}&=& \omega^{\pm}_{0} + \sum_{j=0}^\infty  \alpha^{\pm}_j e^{\pm}_j \,,\label{expansionw}\\
\omega^{s}_{\rm QNM}&=& \omega^{s}_{0} + \sum_{j=0}^\infty  \beta^{s}_j d^s_j \,.\label{expansionwS}
\eeq
The frequencies in GR $\omega^{\pm}_{0},\,\omega^{s}_{0}$ can be found online~\cite{GRITJHU,Leaver:1985ax,Berti:2005ys,Berti:2009kk}.
The corrections in the frequencies are linear in the corrections to the potentials, and the complex numbers
$(e^{\pm}_j,\, d^s_j)$, which represent a ``basis'' set\footnote{We use this terminology for convenience, but we make no claims on the completeness of this basis.} for the perturbations, must be computed numerically. Below, when the context is clear, we omit the superscripts from the coefficients $(e_j,\,d_j)$
for simplicity.

We should note that, generically, there are frequency-dependent terms in the corrections to the potential. 
In our approach, any $\omega$-dependent term in the perturbing potential should be set to $\omega_0$, thereby corresponding to a redefinition of $\alpha^{\pm}_j$. 
We have indeed verified numerically that the induced corrections are linear in the perturbations to the potential, and that $\omega$-dependent terms can be handled by setting them equal to their GR value.

Corrections to the QNM frequencies scale {\it linearly} with corrections to the potential appearing in the master equation. In a forthcoming paper we will show that, when couplings between fluctuations are allowed, corrections in the frequency appear at {\it quadratic} order in the couplings. In some modified gravity theories both corrections are of the same order of magnitude, and therefore couplings cannot be neglected. These issues will be addressed in a separate study~\cite{McManus:2019}.

\begin{figure*}[thbp]
\includegraphics[width=6cm]{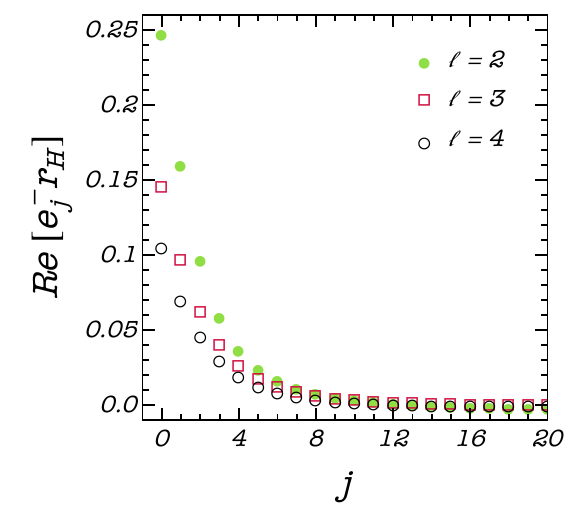}
\includegraphics[width=6cm]{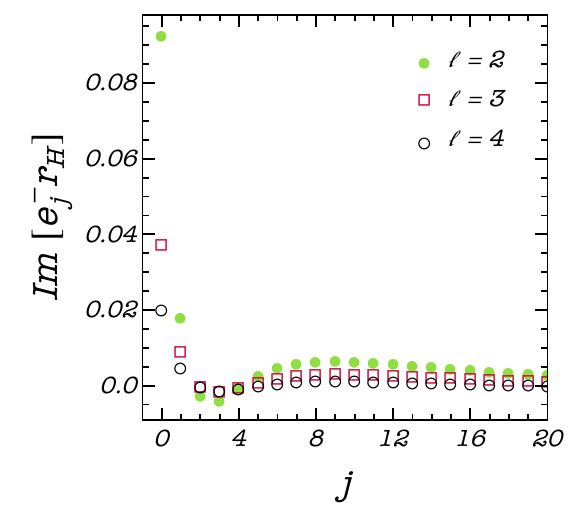}\\
\caption{Real and imaginary parts of the components $e^-_j$ defined in \eqref{expansionw} 
for $j=1,\ldots20$ and odd-parity gravitational perturbations.}
\label{figsomega}
\end{figure*}

\begin{table}[]
\begin{tabular}{ccc}
\hline
\hline
$j$ & $r_H\,e^-_j$            & $r_H\,d^0_j $ \\ 
\hline
\hline
 0 & 0.24725+0.092643i     & 0.15782+0.054078i \\
 1 & 0.15985+0.018208i     & 0.11307+0.015119i  \\
 2 & 0.096632-0.0024155i   & 0.076570+0.00016782i  \\
 3 & 0.058491-0.0037179i   & 0.051121-0.0032973i  \\
 4 & 0.036679-0.00043870i  & 0.034527-0.0024724i  \\
 10 & 0.0036853+0.0065244i & 0.0050350+0.0037363i\\
 \hline
\hline
\end{tabular}
\caption{Real and imaginary part of a few of the first frequency components 
for odd-parity gravitational ($e^-_j$) and scalar field ($d^0_j$) perturbations with $\ell=2$. 
The full set of frequencies up to $j=100$ is provided online~\cite{GRITJHU}.}
\label{table_basis}
\end{table}
%

\section{Results}\label{sec:results}

The most important results of our analysis are the numerical values of the complex factors $e^\pm_j$ and $d^s_j$, which were obtained via high-precision direct integration of the relevant equations.
Figure~\ref{figsomega} and Table~\ref{table_basis} show representative results for the coefficients $e^\pm_j$ and $d^s_j$ in \eqref{expansionw} and \eqref{expansionwS} for odd gravitational and scalar modes, respectively.
Results for all fields and the lowest multipole numbers $\ell=|s|, |s+1| , ..., 10$ are available online~\cite{GRITJHU}.
Our calculated $e_j^\pm$ (and $d_j^s$) have a $5$-digit accuracy
for $j\lesssim 30$.

\subsection{Isospectrality}\label{subsec:iso}
A remarkable result in GR concerns the isospectrality of potentials \eqref{VGRm} and \eqref{VGRplus}. 
These potentials are sometimes referred to as ``superpartners,'' since they can both be obtained from a ``superpotential'' $W_0$~\cite{Chandrasekhar:1985kt,Berti:2009kk}:
\beq
fV_{\pm}&=&W_0^2\mp f\frac{dW_0}{dr}-\frac{\lambda^2(\lambda+2)^2}{36r_H^2}\,,\\
W_0&=&\frac{3r_H(r_H-r)}{r^2(3r_H+\lambda r)}-\frac{\lambda(\lambda+2)}{6r_H}\,.\label{superpotential}
\eeq

When the potentials are perturbed from their GR values as in Eq.~\eqref{expansionV}, isospectrality will in general be broken, unless the coefficients $\alpha^{+}_j$ are related in some way to $\alpha^{-}_j$.

The linearized superpartner relations now yield

\beq
2\frac{d\delta W}{dr}&=&\delta V_{-}-\delta V_{+}\ ,\\
4\frac{W_0}{f}\delta W&=&\delta V_{+}+\delta V_{-}\ ,
\eeq
with $\delta W$ the induced change in the superpotential $W$ with respect to its GR value $W_0$ [Eq.~\eqref{superpotential}].

Let us admit that the theory predicts a set of $\alpha^{-}_j$. The relations above give us a 
unique possibility for $\alpha^{+}_j$, namely:
\beq
\alpha^{+}_0&=&\alpha^{-}_0\,,\quad \alpha^{+}_1=\alpha^{-}_1\,,\\
\alpha^{+}_2&=&\alpha^{-}_2+\frac{6 (\alpha^{-}_0-\alpha^{-}_1)}{\lambda(\lambda+2)}\,,\label{iso_series}
\dots
\eeq
These relations show that isospectrality is very fragile, and that in general it will be broken. 

\section{Asymptotic behavior and convergence of series}\label{sec:asymptotics}
Having defined the ``basis vectors'' $e_j$, it is crucial to investigate their behavior, and the overall convergence 
of the QNM frequency expansion~\eqref{expansionw}. For large $j$, the contribution to the potential from the 
$j$-th term can be described by a Gaussian profile depending on the tortoise coordinate\footnote{The 
Gaussian fit \eqref{gaussian} represents a good approximation for the peak of the potential's modifications $fr_H^2\delta V_j$. 
Although its precision is limited for large values $r_\star$, far from the maximum of the function, the fit is accurate 
enough for the purposes of this section.} $r_\star$, i.e.
\beq
f r_H^2 \delta V_j = f \alpha^{\pm}_j \left(\frac{r_H}{r}\right)^j \simeq \frac{\alpha^{\pm}_j}{e j} e^{-(r_* + r_H \ln j)^2/(2 r_H^2)}\,, \label{gaussian}
\eeq
showing that the location of the peak is proportional to $-\ln j$ (in terms of the tortoise coordinate $r_*$). 
Moreover, since the amplitude of $f \delta V_j$ is proportional to $1/j$ for large $j$, we can expect that $e_j \to 0$ 
for $j \to \infty$. Following \cite{Leung:1999iq}, a perturbative expansion of the QNM frequency yields the 
following correction:
\be
e_j \alpha^{\pm}_j \propto \int_{-\infty}^\infty dr_* f r_H^2 \delta V_j \Phi_0^2\,,\label{omegacorr}
\ee
where $\Phi_0$ is the unperturbed mode itself.

\begin{figure}[t]
\includegraphics[width=4.2cm]{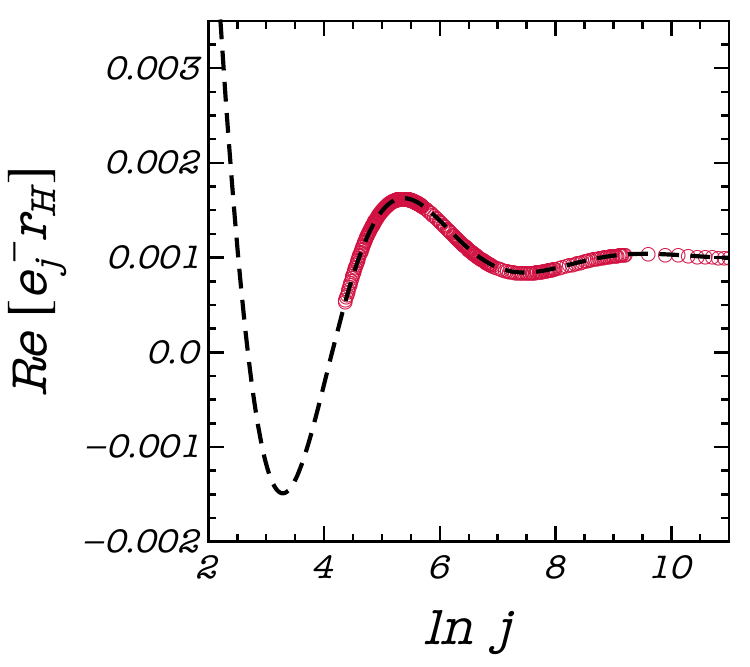}
\includegraphics[width=4.2cm]{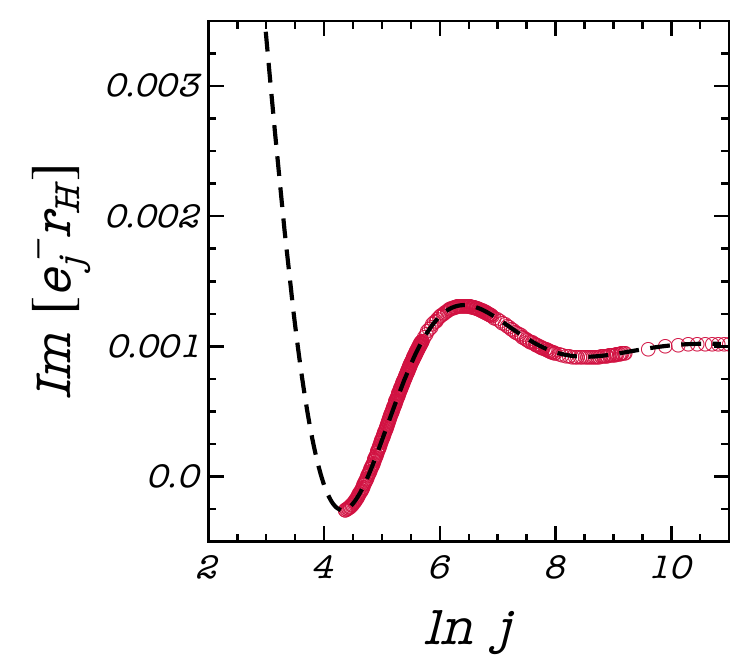}
\caption{Values of the frequency components $e^-_j$ (red dots) for the odd tensor modes with $\ell=2$, compared against the numerical fit of Eq.~\eqref{fitei} (black dashed curve).}
\label{fit}
\end{figure}

Let us assume that $f \delta V_j$ has support only in $-r_H \ln j - r_H \sigma \le r_* \le -r_H \ln j + r_H \sigma$, with a constant $\sigma = {\cal O}(1)$ (the particular choice of $\sigma$ does not affect our results.) Then the integral can be estimated as follows:
\beq
e_j\alpha^{\pm}_j &\propto& \int_{-r_H \ln j - r_H\sigma}^{-r_H \ln j + r_H\sigma} dr_* f \delta V_j \Phi_0^2 \nonumber\\
&=&\int_{-r_H \ln j - r_H \sigma}^{-r_H \ln j + r_H\sigma} dr_* \frac{\alpha^{\pm}_j}{e j} e^{-(r_* + r_H \ln j)^2/(2 r_H^2)} e^{-i 2\omega_0 r_*}
\nonumber\\
&=& \frac{\alpha^{\pm}_j}{j} j^{2ir_H\omega_0} \times {\rm constant}\,.
\eeq
In conclusion we find
\be
e_j \propto \frac{j^{2i\omega_0 r_H}}{j}=
\frac{j^{2 i r_H\omega_R - 2r_H\omega_I}}{j}=\frac{e^{2 i r_H\omega_R \ln j}}{j^{1 +2r_H\omega_I}} \label{eiconv}\,,
\ee
where $\omega_0=\omega_R+i\omega_I$.
Eq.~\eqref{eiconv} shows that the basis frequency $e_j$ decays as $1/j^{1 +2r_H\omega_I}$, and oscillates as $\sin(2r_H\omega_R \ln j)$. For tensor perturbations\footnote{Note that Eq.~\eqref{omegacorr} holds regardless of the overtone number and of the multipole number of the mode, if $j$ is sufficiently large.}
 with $\ell=2$, we have $e_j\simeq\frac{e^{1.5 i \ln j}}{j^{0.64}}$. 
This result is also confirmed by a numerical study of $e_j$ for large $j$. 
We fitted our large-$j$ numerical results to the following functional form:
\begin{equation}
e_{j}\sim\frac{\kappa}{j^\beta}\sin(\gamma \ln j+\zeta)\,,\label{fitei}
\end{equation}
with $(\kappa,\beta,\gamma,\delta)$ numerical coefficients to be determined. We find $\beta \simeq0.66$ and  $\gamma\simeq 1.5$, in very good agreement with the analytical estimate of Eq.~\eqref{eiconv}. Figure~\ref{fit} compares the fit with the actual data for $e_j$.
The agreement is extremely good, and it is strong evidence in support of the asymptotic behavior \eqref{eiconv}.

Having assessed the asymptotic properties of the basis $e_j$, we can now study in detail the convergence of the frequencies 
$\omega_{\rm QNM}$. We assume that the effective potential of the specific theory under consideration is ${\cal C}^\infty$, such that 
we can expand it for large distances as $\delta V=\sum_{j=0}^\infty\frac{a_j}{r^j}$, where the $a_j$'s are constant coefficients. Moreover, 
following our formalism developed in Sec.~\ref{sec:framework}, we expand the corrections with respect to the 
Schwarzschild term according to Eq.~\eqref{expansionV}, such that a mapping exists between the $\alpha^{\pm}_j$ and the coefficients of the theory $a_j$, i.e. $\alpha_j^\pm=r_H^2\frac{a_j}{r_H^j}$.
To study the convergence of the QNM expansion \eqref{expansionw}, we can compute the ratio
\begin{equation}
\Upsilon=\lim_{n\rightarrow\infty}\left\vert\frac{\alpha^\pm_{n+1}e_{n+1}}{\alpha^\pm_ne_n}\right\vert\,.\label{convergence}
\end{equation}
The series converges when $\Upsilon<1$. Replacing the explicit form of the expansion coefficients $\alpha_j^\pm$, 
and using the fact that the frequency ``basis vectors'' $e_j$ behave as in Eq.~\eqref{fitei} -- or equivalently Eq.~\eqref{eiconv} -- 
we find
\beq
\Upsilon=\lim_{n\rightarrow\infty}\left\vert\frac{a_{n+1}}{a_{n}}\frac{n^\beta}{n^{\beta+1}}\frac{\sin[\gamma \ln (n+1)+\zeta]}{r_H\sin[\gamma \ln n+\zeta]}\right\vert=\lim_{n\rightarrow\infty}\left\vert\frac{a_{n+1}}{a_{n}} \right\vert \ .\nonumber
\eeq
The previous equation implies that, in our approach, the convergence of the QNM frequencies depends 
on the behavior of the coefficients of the (expanded) potential for the specific theory of gravity.

As a nontrivial example to test the formalism, consider the case
\be
r_H^2 \delta V = 10^{-2} \frac{\rho_0^k}{\rho_0^k + r^k}=10^{-2} \sum_{j=1}^{\infty} (-1)^{j+1}\left(\frac{\rho_0}{r}\right)^{kj}\,,\label{rho0V}
\ee
where $\rho_0$ is  a constant, and $\alpha_j^\pm = (-1)^{j+1} 10^{-2} (\rho_0/r_H)^{kj}$. For this potential the convergence criterion reduces to 
\begin{equation}
\Upsilon=\lim_{n\rightarrow\infty}\left\vert \frac{\rho_0^{n+1}}{\rho_{0}^n}\right\vert=\vert \rho_0\vert\,,
\end{equation}
and therefore the perturbative formalism is valid if $\rho_0<1$.

\section{Examples}\label{sec:examples}
In this section we will provide some specific examples of theories of
gravity whose gravitational perturbations can be described in terms of
a master equation with the same functional form of
Eq.~\eqref{master}. In order to determine the domain of validity and
the accuracy of our semianalytical approach, we will compare the
results obtained by solving the exact equations of the theory on a
case-by-case basis with the frequencies computed through
Eq.~\eqref{expansionw}.
\subsection{Effective Field Theory}\label{sec:EFT}
An interesting scenario to test against our formalism is given by the BH solutions derived within an EFT approach in Refs.~\cite{Endlich:2017tqa,Cardoso:2018ptl}. Deviations from GR in this theory 
are characterized by three coupling parameters, which define the scale of the strong curvature modifications. For the sake of 
simplicity, we focus on a single coupling parameter, namely 
$\tilde{\Lambda}$ ($\epsilon_2$ in the notation of Ref.~\cite{Cardoso:2018ptl}, which we adopt). In this setup, nonspinning 
BHs are described by the Schwarzschild geometry, and its axial perturbations lead to the following wave equation:
\begin{equation}
\frac{d^2\Psi_-}{dr_\star}+\left[\omega^2-f(V_{-}+\delta V_{-})\right]\Psi_-=0\,,\label{eft}
\end{equation}
where $V_{-}$ is given by Eq.~\eqref{VGRm}, while the extra term in the effective potential reads,
\begin{equation}
\delta V_{-}=\epsilon_2\frac{18(\ell+2)(\ell+1)(\ell-1)r_H^8}{r^{10}}\ .
\end{equation}
Note that in this case $f=1-r_H/r$ and $dr/dr_\star=f$. According the formalism developed in Sec.~\ref{sec:framework}, the only coefficient of the expansion \eqref{expansionw} is then 
\begin{equation}
\alpha^{-}_{10}=18(\ell+2)(\ell+1)(\ell-1)\epsilon_2\ .
\end{equation}

From Table \ref{table_basis} we read off $r_H\omega=r_H[\omega_0+(0.0663354+0.117439 i) \epsilon_2 (\ell-1)(\ell+1)(\ell+2)]$. For $\ell=2$ this leads to a fractional 
deviation from GR of $\epsilon_2(2.13,-15.84)$ in the real and imaginary components, respectively, in excellent agreement with published results: cf. Eqs.~(20) 
of Ref.~\cite{Cardoso:2018ptl}. We find a similar level of agreement for higher-$\ell$ modes.

\subsection{Reissner-Nordstr\"om black holes}\label{sec:rn}
The odd-mode gravitational perturbations of RN BHs are driven by the master equation
\be
f_{\rm RN} \frac{d}{dr}\left(f_{\rm RN} \frac{d\Phi}{dr}\right) + (\omega^2 - f_{\rm RN} V_{\rm RN}) \Phi=0\,,
\ee
with 
\beq
f_{\rm RN}&=& 1 - \frac{2M}{r} + \frac{Q^2}{r^2} = \left(1 - \frac{r_H}{r}\right)\left(1 - \frac{r_-}{r}\right)\,,\\
V_{\rm RN}&=& \frac{\ell(\ell+1)}{r^2}+\frac{4 r_H r_-}{r^4}-\frac{3(r_H+r_-)}{2 r^3}\nonumber\\
&-&\frac{\sqrt{4 (\ell-1) (\ell+2) r_H r_-+\frac{9}{4}(r_H+r_-)^2}}{r^3}\,.
\eeq

We will consider the charge $Q$ -- or, equivalently, the Cauchy horizon radius $r_-\sim Q^2/(2M)$ -- to be a small perturbation.
Note, however, that $r_H$ is {\it not} to be assumed to be equal to $2M$, as it would introduce an error of the same order of the leading order correction.
According to the procedure in Appendix \ref{app:transformation}, 
introducing $\phi  = \sqrt{Z}\Phi = (1-r_-/r)^{1/2}\, \Phi$, we can write the master equation in the form
\be
f\frac{d}{dr}\left(f \frac{d\phi}{dr}\right) +\left[\left(1-\frac{r_-}{r_H}\right)^{-2}\omega^2-f (V_{-} +\delta V)\right]\phi=0\,,
\label{eqrn}
\ee
with
\beq
\delta V =2\frac{r_-}{r_H}\omega^2_0 +\frac{r_-}{r_H^3}\left(\frac{r_H}{r}\right)^3
\left[\frac{5 }{2 }\frac{r_H}{r}-\frac{\lambda+6}{3}\right]\,,
\label{eqrn2}
\eeq
where $f = 1 -r_H/r$ and we ignored ${\cal O}(r_-^2)$ terms.
We can now apply our formalism to RN perturbations.
The coefficients of the effective potential expansion \eqref{expansionw} read: 
\be
\alpha^-_0 =  \frac{2r_-}{r_H} (\omega_0 r_H)^2,
\quad
\alpha^-_3 = -\frac{\left(\lambda+6\right)r_-}{3 r_H},
\quad
\alpha^-_4 = \frac{5 r_-}{2 r_H}\,.\nonumber
\ee
Thus, the QNMs for $\ell = 2$ can be written as
\beq
\omega_{\rm QNM} 
&=&\left(1 - \frac{r_-}{r_H}\right) 
\left(\frac{2 \Omega_0}{r_H} + e_0 \alpha^{-}_0 + e_3 \alpha^{-}_3 + e_4 \alpha^{-}_4\right)\nonumber\\
%
%
%
&=& 
\frac{\Omega_0}{M}
+\frac{(0.0258177\, -0.002824i) Q^2}{M^3}\,,\label{omegaRNqnmM}
\eeq
where $r_H=M+\sqrt{M^2 - Q^2}$, $r_-=M-\sqrt{M^2-Q^2}$, and in the last line we have expanded 
the equation up to order ${\cal O}(Q^2)$. 
Here, $\Omega_0$ is the dimensionless Schwarzschild QNM frequency,
so 
$\Omega_0 = 0.3736716844180418... - i 0.0889623156889357...$ for $\ell =2 $.
\footnote{
Because $\omega_{\rm QNM}$ in our formula is written as a function of $r_H$ and $\alpha_j$,
our formalism requires that one first writes down $\omega_{\rm QNM}$ as a function of $r_H$ and $\alpha_j$.
We note that we should formally use $\omega_0$ in Eq.~\eqref{expansionw} as $2 \Omega_0/r_H$.}
Using the notation of Ref.~\cite{Barausse:2014tra},
these results imply $(\alpha^-_R,\alpha^-_I)=(-6.9,-3.2)\times 10^{-8}[Q/(10^{-3}M)]$, in excellent agreement with reported values in the literature~\cite{Kokkotas:1999bd,Berti:2005eb,Barausse:2014tra}. 

We can quantify the accuracy of our approach and estimate the uncertainty on the non-GR modifications 
of the QNM frequencies by defining
\beq
\Delta_R &=& \left|{\rm Re}(\omega_{\rm QNM}- \omega_{\rm GR})/{\rm Re}(\omega_{\rm full} - \omega_{\rm GR})-1\right|\,,\label{deltaR}
\\
\Delta_I &=& \left|{\rm Im}(\omega_{\rm QNM}- \omega_{\rm GR})/{\rm Im}(\omega_{\rm full} - \omega_{\rm GR})-1\right|\,,\label{deltaI}
\eeq
where $\omega_{\rm full}$ is computed by numerically solving the {\it exact} master equation without any 
approximation, and $\omega_{\rm QNM}$ is given by Eq.~\eqref{omegaRNqnmM}.
For $Q/M = 0.2$, for example, the QNMs frequencies derived using Eq.~\eqref{omegaRNqnmM} agree to within
0.05\%  with those in Ref.~\cite{Kokkotas:1999bd}. The errors obtained for different values of $Q/M$ are listed in Table~\ref{table3}. 
\begin{table}[]
\begin{tabular}{ccc}
\hline
\hline
$Q/M$  & $\Delta_R$& $\Delta_I$\\ 
\hline
$0 .00   $  & $  0  \%    $& $ 0     \%  $\\
$0.05 $ & $  0.11 \% $&$ 0.042 \% $\\
$0.10  $  &$  0.43 \% $&$ 0.17 \% $\\
$0.20  $ &$  1.7 \% $&$ 0.66 \% $\\
$0.30  $  &$  3.8 \% $&$ 1.5 \%  $\\
$0.40  $  &$  6.8 \% $&$ 2.6 \%  $\\
$0.50  $  &$  11 \%  $&$ 4.2  \% $\\
\hline
\hline
\end{tabular}
\caption{Relative percentage errors on the real and imaginary parts of the 
QNMs for RN BHs, as a function of the charge-to-mass ratio $Q/M$.}
\label{table3}
\end{table}

The relative uncertainties $\Delta_{R,I}$ grow with $Q/M$ up to $\sim 10\%$ (4\%) for 
the real (imaginary) component of the mode's frequency when $Q/M=0.5$. This behavior is consistent with 
the assumptions made to obtain Eq.~\eqref{omegaRNqnmM}, in which we neglected terms that are 
order $\mathcal{O}((\alpha^{\pm}_j)^2)$, the latter corresponding to ignoring ${\cal O}(Q^4)$ corrections to the RN perturbations. Therefore, for large values of $Q$, a better accuracy would require to compute $\omega_{\rm QNM}$ by including second-order terms in both $\alpha^{+}_j$ and $\alpha^{-}_j$.
\subsection{Scalars around a slowly-spinning black hole}\label{subsec:Kerr}
The massless Klein-Gordon equation $\square \Psi = 0$ 
around a slowly rotating Kerr BH can be written as
\be
f \frac{d}{dr}\left(f \frac{d}{dr}\right)\Phi + 
\left(\omega^2 - f V_0  - \frac{4 a m M \omega}{r^3}\right)\Phi = 0\,,
\ee
at ${\cal O}(a)$, where $f = 1- 2M/r$, $a/M$ is the BH angular momentum, and
we assumed $\Psi = e^{-i\omega t} Y_{\ell m}(\theta,\phi) \Phi(r)/r$.
We can rewrite this equation in the following form:
\beq
&& f \frac{d}{dr}\left(f \frac{d}{dr}\right)\Phi + 
\bigg[ \left(\omega- \frac{am}{r_H^2}\right)^2
- f \bigg(V_0  
- \frac{2 a m \omega}{r_H^2}
\notag\\ &&
- \frac{2 a m \omega}{r_H^2}\frac{r_H}{r}
- \frac{2 a m \omega}{r_H^2}\left(\frac{r_H}{r}\right)^2
\bigg)\bigg]\Phi = 0,
\eeq
where we used $r_H = 2 M + {\cal O}(a^2)$. Thus
\be
\beta_0^0 = \beta_1^0 = \beta_2^0 = - 2 a m \omega_{0}^0\,,
\ee
and we find:
\be
\omega_{\rm QNM} = \omega_{0}^0 + \frac{am}{r_H^2} - 2 a m \omega_{0}^0 (d_0^0 + d_1^0 + d_2^0)\,.
\ee
Comparing these results with numerical data for $\ell =m=2$~\cite{Berti:2009kk,GRITJHU}, 
we find the relative percentage errors $(\Delta_R, \Delta_I)$ listed in Table~\ref{tablescalar}.
\begin{table}[]
\begin{tabular}{ccc}
\hline
\hline
$a/M$  & $\Delta_R$& $\Delta_I$\\ 
\hline
$0   $&$  0  \%    $& $ 0     \%  $\\
$10^{-4} $&$  0.0050 \% $&$ 0.83 \% $\\
$10^{-3} $&$  0.049 \% $&$ 5.1  \% $\\
$10^{-2} $&$  0.49 \% $&$ 34 \% $\\
\hline
\hline
\end{tabular}
\caption{Relative percentage errors in the real and imaginary parts of
  the QNM frequencies for scalar perturbations around a slowly
  spinning black hole, as a function of the BH angular momentum
  $a/M$.}
\label{tablescalar}
\end{table}
Our formula is a good approximation for very small $a/M \sim 10^{-4}$.
For $a/M \sim 10^{-2}$ the agreement gets worse, especially in the imaginary component.
We find that $a/M \ll 10^{-2}$ is a necessary condition for our approximation to be valid.

\section{Statistical errors}\label{sec:errors}

In this section we analyze the detectability of the modifications of the QNM spectrum by space and terrestrial GW 
interferometers. We follow the approach described in Ref.~\cite{Berti:2005ys}, and we refer the reader to this paper and references 
therein for further details. Since we are interested only in an order-of-magnitude estimate of the associated observational errors, we {\it assume} that the mass of the BH is known (and therefore that the fundamental GR frequencies are known). For a more sophisticated analysis, we refer the reader to Refs.~\cite{Cardoso:2016olt,Meidam:2014jpa}.

The gravitational waveform measured by the interferometers is a linear superposition of 
two polarization states of the form  
$h=h_{+}F_{+}+h_{\times}F_{\times}$,
where $F_{+}$ and $F_{\times}$ denotes the standard pattern functions (which depend on the source orientation with respect 
to the detector and on a ``polarization angle''). 
In the frequency domain, the two GW components are simply given by
\begin{align}
\tilde{h}_+(f)=&\frac{A^+_{\ell mn}}{\sqrt{2}}[e^{i\phi^+_{\ell mn}}S_{\ell mn}b_+(f)+e^{-i\phi^+_{\ell mn}}S^\star_{\ell mn}b_-(f)]\,,\nonumber\\
\tilde{h}_\times(f)=&-\frac{i A^\times_{\ell mn}}{\sqrt{2}}[e^{i\phi^\times_{\ell mn}}S_{\ell mn}b_+(f)+e^{-i\phi^\times_{\ell mn}}S^\star_{\ell mn}b_-(f)]\,,\nonumber
\end{align}
where the amplitude coefficients $A^{+,\times}_{\ell mn}$ and the phase coefficients $\phi^{+,\times}_{\ell mn}$ are real, $S_{\ell mn}$ represent 
the (complex) spin-weighted spheroidal harmonics of spin weight 2, which depend on the polar and azimuthal angles, and $b_{\pm}(f)$ are 
the Breit-Wigner functions: 
\begin{equation}\label{BreitWigner}
b_\pm(f)=\frac{-\textnormal{Im}[\omega_{\ell mn}]}{\textnormal{Im}[\omega_{\ell mn}]^2+(\omega \pm \textnormal{Re}[\omega_{\ell mn}])^2}\ .
\end{equation}
For simplicity, we consider 
$A^\times_{\ell mn}=N_\times A^+_{\ell mn}$ and $\phi^\times_{\ell mn}=\phi^+_{\ell mn}+\phi_0$, where $N_\times$ and 
$\phi_0$ are a scale factor and a phase shift, which are assumed to be known. 

\begin{figure}[t]
\includegraphics[width=8cm]{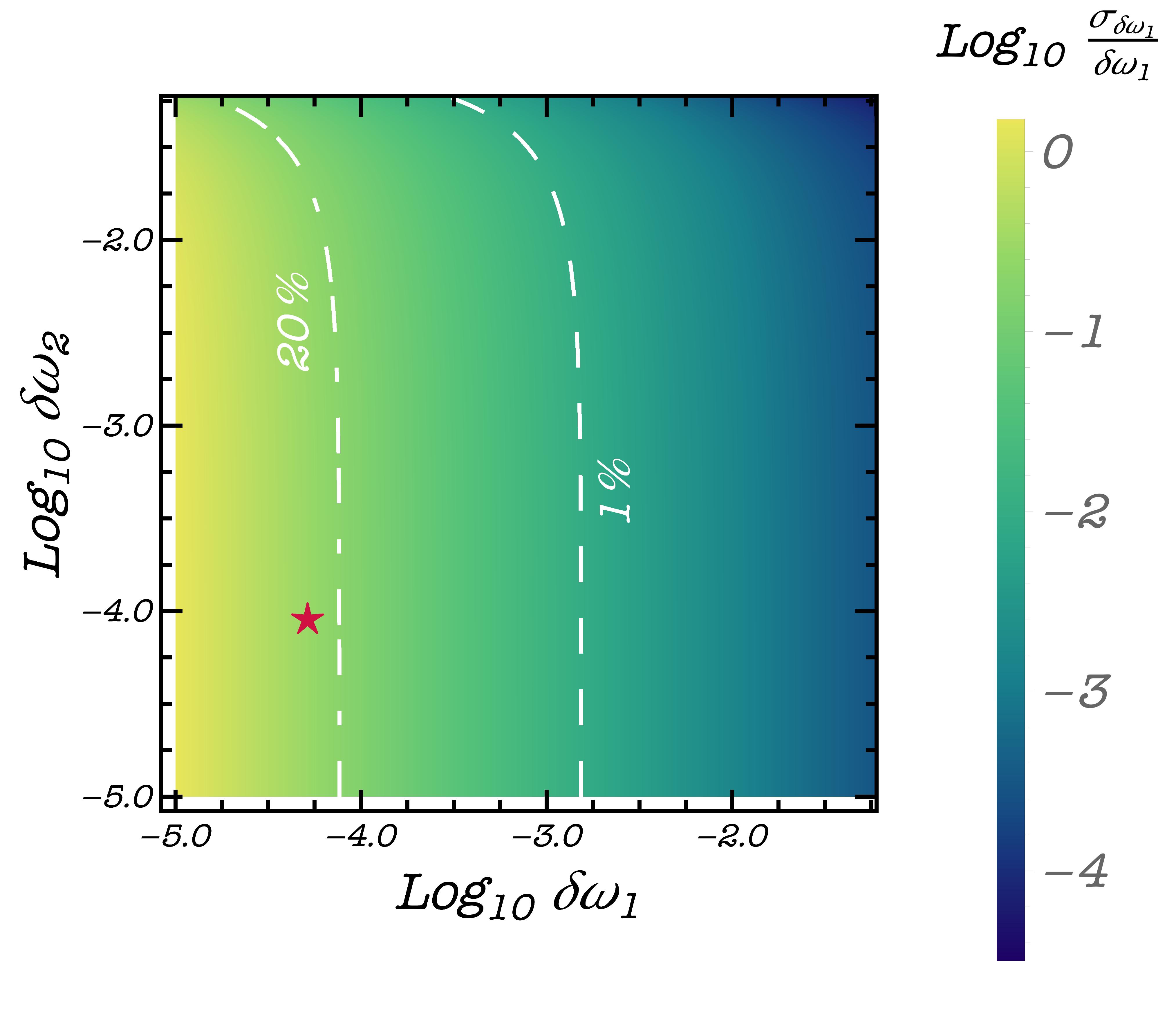}
\caption{Contour plot for the relative errors on the parameters $\delta\omega_1$ which modifies the frequency of the $\ell=2$ fundamental BH mode. Vertical dashed lines represent configurations that lead to 20\% and 1\% fixed uncertainty. The red star identifies the values on $(\delta\omega_1,\delta\omega_2)$ for the EFT model described in Sec.~\ref{sec:EFT}.}
\label{errors}
\end{figure}

Moreover, consistently with the approach introduced in Sec.~\ref{sec:framework}, we express the modes' frequency and damping 
time as (small) deviations of the corresponding Schwarzschild values:
$\omega_{\ell mn}=\omega_{\ell mn}^0+\delta\omega $, where $\delta\omega=\delta\omega_1+i\delta\omega_2$.
The GW signal is then completely specified by 4 parameters, $(A^+_{\ell mn},\phi_{\ell mn}^+,\delta\omega_1,\delta\omega_2)$. 
In order to estimate the errors on the coefficients $\delta\omega_{1,2}$, we consider a Fisher matrix approach, 
which is a reliable approximation of a full Bayesian analysis for signals with high signal-to-noise ratios~\cite{Vallisneri:2007ev}. For this reason we focus 
on possible detections by the space interferometer LISA~\cite{LISA}, an ideal tool to exploit the full potential of QNM 
spectroscopy~\cite{Berti:2007zu,Baibhav:2017jhs,Baibhav:2018rfk}. Finally,
we set $N_\times=1$, and $\phi_{\ell mn}^+=\phi_0=0$. We compute the
errors on the fundamental $\ell=2$ mode of nonspinning BHs, averaging over sky orientation through the following identities:
\begin{align}
\langle F_+^2\rangle=\langle F_\times^2\rangle=\frac{1}{5}\ ,\ \langle F_+F_\times\rangle=0\ \ ,\ 
\langle \vert S_{lmn}\vert^2 \rangle=\frac{1}{4\pi}\ .
\end{align}
Note that the amplitude of the signal depends  in general on the mass, the source 
distance, and the energy released during the ringdown~\cite{Berti:2005ys}. 
For simplicity, here we fix $A^+_{\ell mn}$ by requiring that for $\delta\omega_{1}=\delta\omega_{2}=0$ the signal-to-noise ratio is equal to $10^4$ (a typical value for LISA sources~\cite{Baibhav:2018rfk}). 

The contour plot of Fig.~\ref{errors} shows the relative errors 
$\sigma_{\delta\omega_1}/{\delta\omega_1}$ as a function of various combinations 
of $(\delta\omega_{1},\delta\omega_{2})$. Dashed white curves identify all the possible GR 
modifications that lead to uncertainties of 20\% and 1\%.
We immediately note from the plot that the values of $\sigma_{\delta\omega_1}/{\delta\omega_1}$ are almost completely degenerate with 
respect to $\delta\omega_2$. More interestingly, all the parameter space sampled seems to 
indicate that LISA would be able to provide high-precision measurements of the frequency 
shifts, with a relative accuracy better than $1\%$. This is extremely promising for some of 
the GR modifications which are currently under investigation. For the EFT described in 
Sec.~\ref{sec:EFT}, and identified by a red star in Fig.~\ref{errors}, the frequency change 
$\delta\omega_1$ could be constrained with an error of $\sim 20\%$.

We obtain similar results for the second parameter $\delta\omega_2$, with relative 
uncertainties being always below $\sim 50\%$ for all the configurations analyzed. 
Note finally that a different choice of the waveform parameters, like the quality factor 
$Q_{\ell mn}=\textnormal{Re}[\omega_{\ell m n}]/\textnormal{Im}[\omega_{\ell m n}]/2$, 
may provide better results. We defer more detailed investigations of the 
statistical errors, as well a multiband analysis including third-generation detectors
\cite{Berti:2016lat,Maselli:2017cm}, to future work.

\section{Discussion}\label{sec:discussion}
Our results yield (modulo an algebraic multiplication and sum) the QNM frequencies of any BH spacetime,
provided that it describes a small deviation away from the Schwarzschild geometry. Thus, step 4 in the abstract can be avoided altogether by using our results, which are available online.

There are some important open questions, in particular the extension to coupled master equations~\cite{McManus:2019}. 
The expansion \eqref{expansionV} describes many known theories, and is ``reasonable.'' 
However, our formalism requires $\Upsilon$ in eq.~\eqref{convergence} to be smaller than unity. This restricts the possible space of theories, as we discussed in the example of Eq.~\eqref{rho0V}.
In fact, we suspect that potentials localized far away from the horizon may be hard to accommodate: perturbation theory projects the perturbing potential onto the undisturbed QNMs,
but these diverge exponentially at large distances. Thus, there are known, exponentially sensitive corrections when the potentials have structure at large distances~\cite{Leung:1999iq,Barausse:2014tra,Cardoso:2016rao}.

Alternative expansions are possible, and have been used. Since the GR QNMs are strongly connected to the light ring~\cite{Ferrari:1984zz,Cardoso:2008bp,Berti:2009kk}, a tempting alternative is to expand the perturbing potential around
the light ring~\cite{Glampedakis:2017dvb,Franciolini:2018uyq}. Such an approach is adequate for approximate, WKB-like expansions,
but it may not be suitable for accurate numerical results.

It is intrinsic to our approach (and to many others) that the calculations are performed in a very special coordinate system,
and the gauge invariance of the final results is not obvious. On the other hand, background solutions in generic theories are also usually written in very special coordinate systems. This weakness comes hand in hand with the power of the present approach: it handles any potential, but it is blind to the underlying theory or background. This blindness prevents us from understanding the connection between the expansion parameters in the potentials and the metric multipoles, or to explore the possible connection between geodesic motion and QNMs~\cite{Cardoso:2008bp}.

Finally, we should highlight the fact that our results apply only to those modes which differ slightly from those of GR.
The addition of a small perturbation to a known potential causes the appearance of both a perturbatively small correction to the modes, and new nonperturbative solutions. Take for instance the term $\beta_0$ for scalar perturbations, which is equivalent to a mass parameter $\mu^2=\beta_0/r_H^2$. It is well known that for small $\mu$, the GR QNMs are slightly displaced, {\it and} a new quasi-bound state with frequency $\omega^2\sim\mu^2-r_H^2\mu^4/(4(\ell+1)^2)$ branches off from a zero-mode in the zero-mass limit ~\cite{Detweiler:1980uk,Cardoso:2005vk,Brito:2015oca}. We find, numerically and analytically, that a similar mode exists when $\alpha_0$ is turned on. Such phenomena occur in electromagnetic waveguides as well~\cite{Goldstone:1991pp,Carini:1992dq}. Our analysis is unable to cover systematically the emergence of these new families of modes.

\noindent{\bf{\em Acknowledgments.}}
%
V.C. acknowledges financial support provided under the European Union's H2020 ERC 
Consolidator Grant ``Matter and strong-field gravity: New frontiers in Einstein's 
theory'' grant agreement no. MaGRaTh--646597.
This project has received funding from the European Union's Horizon 2020 research and innovation programme under the Marie Sklodowska-Curie grant agreement No 690904.
AM acknowledges financial support provided under the European Union's H2020 ERC, Starting Grant agreement no. DarkGRA-757480.
We acknowledge financial support provided by FCT/Portugal through grant PTDC/MAT-APL/30043/2017.
We acknowledge the SDSC Comet and TACC Stampede2 clusters through NSF-XSEDE Award Nos. PHY-090003,
as well as MareNostrum and the technical support provided by Barcelona Supercomputing Center (AECT-2018-1-0003).
The authors would like to acknowledge networking support by the GWverse COST Action 
CA16104, ``Black holes, gravitational waves and fundamental physics.''
We acknowledge support from the Amaldi Research Center funded by the 
MIUR program ``Dipartimento di Eccellenza''~(CUP: B81I18001170001).
C.F.B.M. would like to thank the Johns
Hopkins University for kind hospitality during the preparation
of this work and the American Physical Society which funded
the visit through the International Research Travel Award Program.
E.B. and R.M. are supported by NSF Grant No. PHY-1841464, NSF Grant No. AST-1841358, NSF-XSEDE Grant No. PHY-090003, and NASA ATP Grant No. 17-ATP17-0225.
The authors thankfully acknowledge the computer resources, technical expertise and assistance provided by CENTRA/IST. Computations were performed at the cluster “Baltasar-Sete-S\'ois” and supported by the H2020 ERC Consolidator Grant ``Matter and strong field gravity: New frontiers in Einstein's theory'' grant agreement no. MaGRaTh-646597.

\appendix

\section{Klein-Gordon and Maxwell equations in static, spherically symmetric spacetimes}\label{app_wavereduction}
Consider a static, spherically symmetric spacetime
\be
ds^2 = -f_t dt^2 + \frac{1}{f_r}dr^2 + r^2 (d\theta^2  + \sin^2\theta d\phi^2)\,,
\ee
with $f_t=f_t(r), f_r=f_r(r)$. The Klein-Gordon equation $\square \Psi = 0$ on this background spacetime reads
\beq
\sqrt{f_t f_r} \left[\sqrt{f_t f_r} (\Phi^0)' \right]'
+ \left[\omega^2 - f_t \frac{\ell(\ell + 1)}{r^2} - \frac{(f_t f_r)'}{2r}\right]\Phi^0 = 0\,,\nonumber
\eeq
where primes stand for radial derivatives, and the scalar $\Psi = e^{-i\omega t}Y_{\ell m}\Phi^0(r)/r$.
Similarly, the Maxwell equations $\nabla_\mu F^{\mu \nu} = 0$ with $F_{\mu \nu} = \partial_\mu A_\nu - \partial_\nu A_\mu$
reduce to a single master equation
\be
\sqrt{f_t f_r} \left[\sqrt{f_t f_r} (\Phi^{1,\pm})' \right]' + 
\left[\omega^2 - f_t \frac{\ell(\ell + 1)}{r^2} \right]\Phi^{1,\pm} = 0\,,\nonumber
\ee
where we assumed 
\beq
A_\mu dx^\mu &=& a_t(r) e^{-i\omega t} Y_{\ell m} dt + a_r(r) e^{-i\omega t} Y_{\ell m} dr 
\notag\\ && + a_+(r)e^{-i\omega t} (\partial_\theta Y_{\ell m}d\theta + \partial_\phi Y_{\ell m}d\phi)
\notag\\ && 
+ a_-(r)e^{-i\omega t} \left( \frac{\partial_\phi Y_{\ell m}}{\sin\theta} d\theta - \partial_\theta Y_{\ell m}d\phi\right)\,,
\label{generalmaxwelleq}
\eeq 
and 
\be
\Phi^{1,+} = \frac{\sqrt{f_t f_r} a_r}{i\omega}, \quad \Phi^{1,-} = a_-\,.
\ee
When the spacetime is close to the Schwarzschild spacetime, we can assume $f_t = (1 - r_H/r)[1 + \delta_t(r)], f_r = (1 - r_H/r)[1 + \delta_r(r)]$, with $\delta_{t,r}$ small quantities. Therefore the perturbations induced by scalar and vector fields on a spherically symmetric background can always be cast in the form
\be
F \frac{d}{dr}\left[F \frac{d\Phi}{dr} \right]+(\omega^2 - F \bar{V})\Phi = 0,
\label{generalmastereq}
\ee
with
\be
F(r)\equiv f(r) Z(r) = \left(1 - \frac{r_H}{r}\right)  Z(r).
\ee

The description of tensor perturbations requires a knowledge of the underlying gravitational theory. While we do not have a rigorous proof, we expect that under the assumptions spelled out in the Introduction it should be possible to reduce tensor perturbations to the study of Eq.~\eqref{generalmastereq}. As shown in Ref.~\cite{Cardoso:2018ptl}, even when the equations of motion are of higher order, the relations between metric quantities in GR at zeroth order can be used to eliminate higher-order derivatives. Then, algorithms similar to that introduced within GR in Ref.~\cite{Zerilli:1970se} can be used to cast the equations in the form~\eqref{generalmastereq}.

\section{Transformation to Eq. (1)}\label{app:transformation}
As shown in Appendix~\ref{app_wavereduction}, when working out the perturbation equations the ``natural'' form of the master equation is not our Eq.~\eqref{master}, but rather Eq.~\eqref{generalmastereq}.
Defining $\phi \equiv \sqrt{Z} \Phi$, Eq.~\eqref{generalmastereq} reads
\beq
\label{eq:intermediate}
&&f \frac{d}{dr}\left[f \frac{d\phi}{dr} \right]+\left[\frac{\omega^2}{Z^2} - f V\right]\phi=0\,,\\
&&V = \frac{\bar{V}}{Z} - \frac{f (Z^\prime)^2 - 2 Z (f Z^\prime)^\prime}{4 Z^2}\,.
\eeq
If $\bar{V}$ and $Z$ are written in terms of small perturbations $\delta \bar{V}$ and $\delta Z$ as
\beq
\bar{V} &=& V_{\rm GR} + \delta \bar{V}\,,\\
Z &=& 1 + \delta Z\,,
\eeq
then the potential $V$ becomes (to leading order in the perturbations)
\be
V = V_{\rm GR} + \delta \bar{V}  - V_{\rm GR} \delta Z + \frac{1}{2}(f \delta Z^\prime)^\prime \,,
\ee
and the frequency-dependent term in Eq.~(\ref{eq:intermediate}) reads
\beq
\frac{\omega^2}{Z^2} &=&\omega^2 \left[1 -2 \delta Z(r_H)\right] - 2 \omega^2 \left[\delta Z(r) - \delta Z(r_H)\right]\,.
%
\eeq
The factor $\left[1 -2 \delta Z(r_H)\right]$ in the first term on the right-hand side amounts to a rescaling of the frequencies.
The second term on the right-hand side vanishes at $r=r_H$, so one can factor out $f$ in this term and absorb it in the perturbed potential. Therefore Eq.~\eqref{generalmastereq} is equivalent to the starting point of our analysis, i.e. Eq.~\eqref{master}.

\section{$e_2^\pm$ and $e_3^\pm$ for large $\ell$}\label{app:eikonal}
We can regard the deviation between $V_+$ in Eq.~\eqref{VGRplus} as $V_-$ in Eq.~\eqref{VGRm}
as a small correction
 for large $\ell$, because $V_+ - V_- = {\cal O}(\ell^{-2})$.
More explicitly, we have
\beq 
V_+ &=& V_- 
+ \frac{1}{r_H^2}\left[- \frac{12}{\ell^2} \left(\frac{r_H}{r}\right)^3
+
\frac{18}{\ell^2} \left(\frac{r_H}{r}\right)^4
\right] 
\notag\\ &+& 
\frac{1}{r_H^2}\left[ \frac{12}{\ell^3} \left(\frac{r_H}{r}\right)^3
-
\frac{18}{\ell^3} \left(\frac{r_H}{r}\right)^4
\right]
+{\cal O}(\ell^{-4}).
\eeq
Since $V_+ - V_-$ can be expanded in powers of $r_H/r$ and the series is convergent in $r_H \le r < \infty$,
we can apply our formalism to this case.
From our QNM formula, the relation
\beq
\omega_{\rm QNM}^+ &=& 
\omega_{\rm QNM}^-  + 6(-2 e_2^- + 3 e_3^-)\left(\frac{1}{\ell^2} - \frac{1}{\ell^3} \right)
+ {\cal O}(\ell^{-4}),
\notag
\eeq
should hold. 
On the other hand, because QNM frequencies for even and odd modes are the same, we can obtain the large-$\ell$ relations
\beq
e_3^- &=& 2 e_2^- /3 + {\cal O}(\ell^{-2})\,,\\
e_3^+ &=& 2 e_2^+ / 3 + {\cal O}(\ell^{-2})\,.
\eeq
A numerical calculation shows that
$|1- 2 e_2^- /(3 e_3^-)|$ is
$0.016$ for $\ell = 10$,
and $0.0082$ for $\ell = 20$. 
This is consistent with the fact that the error is of order ${\cal O}(\ell^{-2})$.
We should note that since our QNM formula
contains an error of order ${\cal O}((\alpha_2)^2) = {\cal O}(\ell^{-4})$, 
we cannot obtain relations for $e_j$ with $j\ge 4$.

\bibliography{QNMTaylor_refs}

\end{document}